\documentclass[journal,twoside,web]{ieeecolor}
\usepackage{tmi}

\usepackage{amsmath,amsfonts,amssymb}
\usepackage{graphicx}
\usepackage{xcolor}
\usepackage{multicol}

\usepackage{url}

\usepackage{algorithm}
\usepackage{algorithmic}

\usepackage{tikz}

\usepackage{pgfplots}
\usepackage{filecontents}
\pgfplotsset{compat=1.5.1}

\tikzset{
  pics/square/.default={1},
  pics/square/.style = {
    code = {
    \draw[pic actions] (0,0) rectangle (#1,0.6#1);
    }
  }
}

\usepackage{nicefrac}

\usepackage{bbm}
\renewcommand{\epsilon}{\varepsilon}				% nicer Epsilon
\renewcommand{\theta}{\vartheta}                  	% nicer Theta
\renewcommand{\phi}{\varphi}						% nicer Phi

\newcommand{\zb}[1]{\mbox{\boldmath{${#1}$}}}
\newcommand{\zbs}[1]{\mbox{\boldmath\scriptsize{${#1}$}}}

\newcommand{\adj}{{\vdash \hspace*{-1.752mm} \dashv}}

\usepackage[binary-units = true,exponent-product = \cdot]{siunitx}
\DeclareSIUnit{\vacuumpermeability}{\text{$\mu_0$}}

\pgfplotsset{
colormap/inferno/.style={%
    /pgfplots/colormap={inferno}{%
      rgb=(0.001462, 0.000466, 0.013866)
      rgb=(0.037668, 0.025921, 0.132232)
      rgb=(0.116656, 0.047574, 0.272321)
      rgb=(0.217949, 0.036615, 0.383522)
      rgb=(0.316282, 0.053490, 0.425116)
      rgb=(0.410113, 0.087896, 0.433098)
      rgb=(0.503493, 0.121575, 0.423356)
      rgb=(0.596940, 0.154848, 0.398125)
      rgb=(0.688653, 0.192239, 0.357603)
      rgb=(0.775059, 0.239667, 0.303526)
      rgb=(0.851384, 0.302260, 0.239636)
      rgb=(0.912966, 0.381636, 0.169755)
      rgb=(0.956852, 0.475356, 0.094695)
      rgb=(0.981895, 0.579392, 0.026250)
      rgb=(0.987464, 0.690366, 0.079990)
      rgb=(0.973088, 0.805409, 0.216877)
      rgb=(0.947594, 0.917399, 0.410665)
      rgb=(0.988362, 0.998364, 0.644924)
    },
  },}

\definecolor{ibilight}{RGB}{193,216,237}
\definecolor{ibidark}{cmyk}{1.00,0.69,0.00,0.12}
\definecolor{uke2}{cmyk}{0.10,0.18,0.25,0.36}
\definecolor{uke3}{cmyk}{0.00,0.00,0.00,0.80}
\definecolor{ukesec1}{cmyk}{0.00,0.09,1.00,0.00}
\definecolor{ukesec2}{cmyk}{0.00,0.61,0.99,0.00}
\definecolor{ukesec3}{cmyk}{0.59,0.00,0.22,0.00}
\definecolor{ukesec4}{cmyk}{0.54,0.00,1.00,0.00}
\definecolor{tuhh}{cmyk}{0.65,0.00,0.20,0.00}

\usetikzlibrary{positioning}
\usepgfplotslibrary{units}

\usepackage{flushend}

\begin{document}

\title{Efficient Joint Estimation of Tracer Distribution and Background Signals in Magnetic Particle Imaging using a Dictionary Approach}

\author{Tobias~Knopp, Mirco~Grosser, Matthias~Graeser,  Timo~Gerkmann, Martin~M{\"o}ddel
\thanks{This work was supported in part by the German Research Foundation (DFG) under Grant KN 1108/2-1 and in part by the Federal Ministry of Education and Research (BMBF) under Grant 05M16GKA and Grant 13XP5060B.}
\thanks{T. Knopp, M. Grosser, M. Graeser, and M. M{\"o}ddel are with the Section for Biomedical Imaging, University Medical Center Hamburg-Eppendorf, 20246 Hamburg, Germany and the Institute for Biomedical Imaging, Hamburg University of Technology, 21073 Hamburg, Germany (e-mail: t.knopp@uke.de). T. Gerkmann is with the Signal Processing Group, University Hamburg, 22527 Hamburg}
%\thanks{Copyright (c) 2021 IEEE. Personal use of this material is permitted. However, permission to use this material for any other purposes must be obtained from the IEEE by sending a request to pubs-permissions@ieee.org.}
} % <-this % stops a space

% make the title area
%\vspace{-80pt}
\maketitle

% As a general rule, do not put math, special symbols or citations
% in the abstract or keywords.

\begin{abstract}
Background signals are a primary source of artifacts in magnetic particle imaging and limit the sensitivity of the method since background signals are often not precisely known and vary over time. The state-of-the art method for handling background signals uses one or several background calibration measurements with an empty scanner bore and subtracts a linear combination of these background measurements from the actual particle measurement. This approach yields satisfying results in case that the background measurements are taken in close proximity to the particle measurement and when the background signal drifts linearly. In this work, we propose a joint estimation of particle distribution and background signal based on a dictionary that is capable of representing typical background signals. Reconstruction is performed frame-by-frame with minimal assumptions on the temporal evolution of background signals. Thus, even non-linear temporal evolution of the latter can be captured. Using a singular-value decomposition, the dictionary is derived from a large number of background calibration scans that do not need to be recorded in close proximity to the particle measurement. The dictionary is sufficiently expressive and represented by its principle components. The proposed joint estimation of particle distribution and background signal is expressed as a linear Tikhonov-regularized least squares problem, which can be efficiently solved. In phantom experiments it is shown that the method strongly suppresses background artifacts and even allows to estimate and remove the direct feed-through of the excitation field.
\end{abstract}

% Note that keywords are not normally used for peerreview papers.

%\IEEEpeerreviewmaketitle

\section{Introduction}
\IEEEPARstart{T}{omographic} imaging is one of the most important tools for making a diagnosis in modern medicine. Even though imaging modalities like computed tomography and magnetic resonance imaging (MRI) do not require tracer material for generating an image of the human body, in clinical practice tracers are used nevertheless since they enhance the contrast of the images. Super-paramagnetic iron-oxide (SPIO) nanoparticles are well researched tracers \cite{bauer2015magnetic} that are bio-compatible and therefore suitable for tomographic imaging. In MRI they are not in wide-spread use since they generate negative instead of positive contrast. To circumvent these drawbacks of SPIO imaging with MRI the researchers B. Gleich and J. Weizenecker developed an entirely new tomographic imaging technique named magnetic particle imaging (MPI)~\cite{Gleich2005Nature,knopp2017magnetic} that allows to image SPIOs with positive contrast and without any tissue-background signal. While MPI is still in an early stage of development, it has already shown in pre-clinical settings to be suitable for the detection of stroke \cite{ludewig2017magnetic}, bleeding \cite{yu2017magneticB,szwargulski2020monitoring}, cancer \cite{arami2017tomographic}, stenoses \cite{vaalma2017magnetic}, and the presence of cerebral aneurysms \cite{sedlacik2016magnetic}.  Furthermore, MPI has proven to be suited for visualizing lung perfusion \cite{zhou2017first}, labeled stem cells \cite{bulte2015quantitative}, and cerebral blood volume \cite{cooley2018rodent}. A route for first human application of MPI was sketched with the first human-scale brain imager~\cite{graser2019human}.

The key to obtaining high quality MPI tomograms of high spatial resolution and high signal-to-noise ratio (SNR) is to have a very sensitive imaging system, which mainly detects the signal generated by the tracer. In theory, the sensitivity is limited only by noise in the electronic components of the receive chains, which can be assumed to follow a Gaussian statistic. In practice, however, additional perturbations are present. For instance, thermal effects in the scanner can lead to a slight drift of the induced signal. The limitation in sensitivity was first investigated in~\cite{them2016sensitivity}. As a solution, the authors proposed to take an empty measurement with a free scanner bore prior to the actual measurement and subtract the empty measurement prior to reconstruction, which enhanced the sensitivity by more than one order of magnitude.

The static background subtraction works well as long as the background signal remains static over time. For changing background signals it was proposed in~\cite{knopp2019correction} to use two background measurements, one directly before and one directly after the experiment. By using a convex combination of both scans it is possible to significantly reduce artifacts induced by dynamic background signals. Still, this method has two limitations: First, it can only correct linear changes of the background signal, which limits the method to short measurement scenarios. Second, it complicates the entire measurement protocol since the measured subject needs to be pulled in and out quickly before and after the actual experiment.

An alternative to this approach is to estimate the background signal directly from the measured data, which was proposed in~\cite{straub2018joint}. To this end, the authors proposed a joint estimation of tracer distribution (foreground signal) and background signal. Since this general optimization approach has no unique solution, it was proposed to shift the field-of-view (FoV) slightly from frame to frame using a tailored multi-patch sequence %\cite{Knopp2015PhysMedBiol} 
and assume that the foreground and the background did not change in-between two frames. This puts a constraint on the optimization problem leading to a unique solution. We note, however, that the method decreases the temporal resolution by at least a factor of two and requires a very special measurement protocol that only few MPI scanners are capable of.

Since both the linear interpolation approach in~\cite{knopp2019correction} and the joint estimation approach in~\cite{straub2018joint} require an advanced measurement protocol, the purpose of the present paper is the development of a method that can be applied to any MPI measurement sequence and scanner and does not alter the imaging protocol. One further requirement is that the reconstruction time should not be substantially increased. Here, we note that the simple background subtraction in~\cite{knopp2019correction} does not increase the reconstruction time whereas the approach in~\cite{straub2018joint} in its published form leads to an increase in reconstruction time effectively preventing real-time reconstruction~\cite{knopp2016online,vogel2017low}. 

Our approach uses a dictionary for representing the low-dimensional subspace containing all typical background signals. By constraining the background signal to be part of this subspace and by forcing the particle signal to follow the MPI signal model it is possible to estimate both quantities in a joint fashion and with no increase of the algorithmic complexity. Our approach has similarities to a background estimation discussed in~\cite{von2017hybrid} where an additional background pixel was used for correcting a linear scaling of a single background measurement. It can be seen as a generalization of that method and we show that a multi-dimensional space of background signals reduces the artifacts.

\section{Theory}

We use typical mathematical notation with boldface letters for vectors and matrices. The identity matrix of size $N\times N$ is denoted by $\zb I_N$. Zero vectors and matrices are written as $\zb 0$ where the size is not explicitly mentioned but can be derived from the context. For a vector $\zb x =( x_n )_{n=1}^N \in \mathbb{C}^N$ we define a projection operator $P_{j,k}:\mathbb{C}^N\rightarrow\mathbb{C}^{k-j+1}$, $1\le j\le k\le N$ by
\begin{align*}
    %\zb x[j:k] & := ( x_n )_{n=j}^k  \in \mathbb{C}^{k-j+1} \\
    P_{j,k}(\zb x) & := ( x_n )_{n=j}^k%\\
\end{align*}
that outputs a sub-vector of $\zb x$. Analogously, for a matrix $$\zb A=( A_{n,r} )_{n=1,\dots,N;r=1,\dots,M} \in \mathbb{C}^{N\times M}$$ we define a projection operator $P_{j,k,l,m}:\mathbb{C}^{N\times M}\rightarrow\mathbb{C}^{(k-j+1)\times (m-l+1)}$, $1\le j\le k\le N$, $1\le l\le m\le M$ by
\begin{align*}
    %\zb x[j:k] & := ( x_n )_{n=j}^k  \in \mathbb{C}^{k-j+1} \\
    P_{j,k,l,m}(\zb A) & := ( A_{n,r} )_{n=j,\dots,k; r=l,\dots,m} %\\
\end{align*}
that outputs a sub-matrix at the defined index ranges.

\subsection{Ideal Imaging Equation}

We consider a typical MPI experiment where the tracer distribution is periodically excited and a sequence of $L$ frames is continuously measured. The voltage signals induced in one or multiple receive coils are Fourier transformed frame by frame and in an optional step a frequency filtering is applied. We let $\zb u_l^\text{ideal} = \left( \hat{u}_{m,l} \right)_{m=1,\dots,M} \in \mathbb{C}^{M}$ denote the ideal  background-free measurement vectors that are generated by the tracer distribution $\zb c_l \in \mathbb{R}_+^{N}$ where $\mathbb{R}_+$ are the positive real numbers including zero. In the remainder of this manuscript we omit the frame index and write $\zb c$ and $\zb u^\text{ideal}$ when a fixed frame is considered and the frame dependency is not important. The relation between $\zb u^\text{ideal}$ and $\zb c$ is linear and can be expressed as
\begin{equation} \label{Eq:LSIdeal}
    \zb u^\text{ideal}=\zb S \zb c
\end{equation}
where $\zb S \in \mathbb{C}^{M\times N}$ is the MPI system matrix.

\subsection{Background Signals}

In practice, it is not possible to obtain the idealized signal $\zb u^\text{ideal}$ directly since one instead measures a noisy measurement $\zb u_l$. The latter can be described as the superposition of three components:
\begin{equation}
    \zb u_l = \zb u_l^\text{ideal} + \underbrace{\zb b^\text{static} + \zb b_l^\text{dynamic}}_{=:\zbs b_l},
\end{equation}
where $\zb b^\text{static} \in\mathbb{C}^{M}$ is a background signal that remains static independently of the frame index $l$, and $\zb b_l^\text{dynamic}\in\mathbb{C}^{M}$ is a dynamic background signal that changes over time.

\subsection{A-priori Background Correction}

Since the background signal is superimposing the idealized signal it is possible to remove it by subtraction. Let us assume that $\zb b^\text{est}$ is an estimate of the background signal. Then, one can correct the measured signal $\zb u$ by calculating
\begin{equation} \label{Eq:RecoBGSubst}
    \zb u^\text{corr} = \zb u - \zb b^\text{est}
\end{equation}
One way to obtain an estimator for the background signal is to directly measure the background signal $\zb b_\text{static}$ by removing all tracer from the scanner $\zb c^\text{empty} = \zb 0$ so that $\zb u^\text{empty} - \zb b = \zb 0$ and hence $\zb u^\text{empty} = \zb b$. Here, one has to keep in mind that the background measurement is linked to the specific time point when $\zb u^\text{empty}$ was measured, which is different from the time point when $\zb u$ was measured. This is why the dynamic part of the background signal $\zb u^\text{empty}$ will be different from the one in $\zb u$. 

It is therefore advantageous to get rid of the dynamic part in the empty measurement. This can be achieved quite easily, if the expectation value of the dynamic part vanishes $E(\zb b^\text{dynamic})=\zb 0$, by obtaining a large number of samples for the background signal such that $E(\zb u^\text{empty}) \approx \zb b^\text{static}$, where $E(\cdot)$ denotes the statistical expectation operator. 

In case the expectation value of the dynamic part does not vanish, e.g. due to drifts in the dynamic background signal, different strategies are required. One way is to measure two background measurements, one before  ($\zb u_\text{pre}$) and one after  ($\zb u_\text{post}$) the experiment. Then, the $l$-th background signal $\zb b_l$ can be approximated by the convex combination
\begin{equation*}
 \zb u^\text{dyn}_l = \frac{L-l}{L-1} \zb u_\text{pre} + \frac{l-1}{L-1} \zb u_\text{post}
\end{equation*}
and subtracting it in the same way as the static background measurement in \eqref{Eq:RecoBGSubst}. While this approach performs well in some applications~\cite{knopp2019correction} it has some clear limitations:
\begin{itemize}
    \item It only works when the drift is at least approximately linear. Typically, this holds true only for short measurement sequences with a small number of measured frames.
    \item It requires background measurements taken in close temporal proximity to the measurement, which complicates the measurement protocol.
\end{itemize}
Throughout this work the subtraction of a linear interpolated background signal will be considered as the state of the art reference method for background signal estimation.

\subsection{Standard Image Reconstruction}

The standard approach to reconstruct the tracer distribution $\zb c$ from the measurements $\zb u$ is to solve the regularized least-squares problem
\begin{equation} \label{Eq:RecoStd}
 \underset{\zbs c}{\text{argmin}}  \Vert \zb S \zb c - \zb u + \zb b^\text{est} \Vert_2^2 + \lambda \Vert \zb c \Vert_2^2.
\end{equation}
where the background estimate $\zb b^\text{est}$ is subtracted from the measurement in the data discrepancy term. Regularization is required since the corrected measurements still contain a noise component, which will be amplified by the ill-conditioned MPI system matrix without regularization.

\subsection{Joint Estimation using a Background Dictionary} \label{sec:BGDict}

The standard approach \eqref{Eq:RecoBGSubst} requires an accurate estimate of the background signal $\zb b$. In case that the estimate is poor, e.g. because the background is drifting over time, the standard approach may lead to image artifacts since parts of the background signal are reconstructed into image space. This motivates us to investigate an adaptive \textit{joint estimation} of the tracer distribution $\zb c$ and background signal $\zb b$, which operates frame-by-frame and relaxes the assumptions on stationarity or linearity of the temporal evolution of background signals. In the most general form, this can be formulated as 
\begin{equation} \label{Eq:RecoBGNaiv1}
 \underset{\zbs c, \zbs b}{\text{argmin}}  \Vert \zb S \zb c - \zb u + \zb b \Vert_2^2 +  R^\text{fg}_\lambda(\zb c) + R^\text{bg}_\beta(\zb b)
\end{equation}
where $R^\text{fg}_\lambda(\zb c)$ is a regularization term constraining the particle concentration $\zb c$, which is chosen to be $R^\text{fg}_\lambda(\zb c) = \lambda \Vert \zb c\Vert_2^2$ in this work. $R^\text{bg}_\beta(\zb b)$ is a regularization term constraining the background signal. Without this additional regularization term, the optimization problem \eqref{Eq:RecoBGNaiv1} would have the trivial solution $\zb c = \zb 0$, $\zb b =-\zb u$, which fails to provide any useful information on the particle distribution.

The core proposal of the paper is to use an orthogonal dictionary $\zb \Phi \in \mathbb{C}^{M \times Q}$ providing a basis for the space of background signals. Let 
\begin{equation}
    \Gamma := \{ \zb b \in \mathbb{C}^M\; | \; \zb \Phi \zb n = \zb b\; \text{where} \; \zb n \in \mathbb{C}^Q \}
\end{equation}
be the subspace spanned by the dictionary $\zb \Phi$.

Then, the proposed constraint can be incorporated by using a regularization term of the form
\begin{equation} \label{Eq:BGRegularizer}
  R^\text{bg}_\beta(\zb b) := \tilde{R}^\text{bg}_\beta( \zb b - \zb  b^\text{est}) + \chi_{\Gamma}(\zb b - \zb  b^\text{est})
\end{equation}
where $\chi_{\Gamma}$ is the indicator function
\begin{equation*}
   \chi_{\Gamma}(\zb b) := \begin{cases} 0 & \zb b \in \Gamma \\
   \infty & \text{else} \\
                            \end{cases}.  
\end{equation*}
and the term $\zb  b^\text{est}$ is included as an initial guess so that $\zb b$ will only include differences to the static background.
While the indicator function ensures that $\zb b - \zb b^\text{est}$ is contained in the background space $\Gamma$, the first term, originally proposed in~\cite{straub2018joint}, should ensure that $\zb b$ is similar to the estimate $\zb  b^\text{est}$.

Before, further specifying the form of $\tilde{R}^\text{bg}_\beta$, we bring problem \eqref{Eq:RecoBGNaiv1} with penalty \eqref{Eq:BGRegularizer} into a form more suitable for its solution. To this end, we first apply a substitution $\zb b \mapsto \zb b + \zb  b^\text{est}$ to move $\zb  b^\text{est}$ into the data discrepancy term yielding
\begin{equation} \label{Eq:17}
 \underset{\zbs c, \zbs b}{\text{argmin}}  \Vert \zb S \zb c - \zb u + \zb b + \zb  b^\text{est} \Vert_2^2 + \lambda \Vert \zb c \Vert_2^2 + \tilde{R}^\text{bg}_\beta( \zb b ) + \chi_{\Gamma}(\zb b).
\end{equation}
Then, we replace the optimization variable $\zb b$ by $\zb \Phi \zb n$, which ensures that $\chi_{\Gamma}(\zb \Phi \zb n) =  0$. Therefore \eqref{Eq:17} can be reformulated as
\begin{equation} \label{Eq:Deriv2}
 \underset{\zbs c, \zbs n}{\text{argmin}}  \Vert \zb S \zb c - \zb u + \zb \Phi \zb n + \zb  b^\text{est} \Vert_2^2 + \lambda \Vert \zb c \Vert_2^2 + \tilde{R}^\text{bg}_\beta( \zb n ),
\end{equation}
where $\zb n$ is the new optimization variable for the background. Moreover, we note that the original regularization $\tilde{R}^\text{bg}_\beta( \zb b )$ term can be reformulated in terms of $\zb n$ in a straight forward manner. In order to make sure that the estimated background signal remains similar to the initial estimate, we regularize $\zb n$ using a weighted $\ell_2$-norm $\tilde{R}^\text{bg}_\beta( \zb n ) = \beta\Vert \zb W^{\frac{1}{2}}\zb n \Vert^2$. Thus our inverse problem becomes
\begin{equation}\label{Eq:RecoBGNFull}
 \underset{\zbs c, \zbs n}{\text{argmin}}  \Vert \zb S \zb c - \zb u + \zb \Phi \zb n + \zb  b^\text{est} \Vert_2^2 + \lambda \Vert \zb c \Vert_2^2 + \beta\Vert \zb W^{\frac{1}{2}}\zb n \Vert^2.
\end{equation}
In the simplest case one could chose $\zb W = \zb I_Q$, which is equivalent to applying classical Tikhonov regularization to the background signal, i.e. $\tilde{R}^\text{bg}_\beta( \zb b ) = \Vert \zb b \Vert_2^2$. However, the weighting matrix $\zb W$ can also be used to incorporate further prior knowledge about the background signals at hand. Our choice for $\zb W$ will be discussed in more detail in Sec. \ref{sec:SetupDict}.

To efficiently, solve problem \eqref{Eq:RecoBGNFull}, we first define
\begin{align*}
    \zb D & := \begin{pmatrix} \lambda^{\frac{1}{2}} \zb I_N & \zb 0 \\ \zb 0 & \beta^{\frac{1}{2}} \zb W^{\frac{1}{2}}
    \end{pmatrix}, \quad \zb y := \begin{pmatrix} \zb c \\ \zb n \end{pmatrix}.
\end{align*}
Now we can pull the regularization parameters into the norms, and stack $\Vert \lambda^{\frac{1}{2}} \zb c \Vert_2^2$ and $\Vert \beta^{\frac{1}{2}}  \zb W^{\frac{1}{2}} \zb n \Vert_2^2$ together yielding
\begin{equation} \label{Eq:RecoBGNaiv3}
 \underset{\zbs c, \zbs n}{\text{argmin}}  \Vert \zb S \zb c - \zb u + \zb \Phi \zb n  + \zb  b^\text{est} \Vert_2^2 + \Vert\zb D \zb y \Vert_2^2,  
\end{equation}
Finally, we define
\begin{align*}
    \zb A & := \begin{pmatrix} \zb S & \zb \Phi\end{pmatrix}, \quad \zb w := \zb u - \zb  b^\text{est} 
\end{align*}
and can express \eqref{Eq:RecoBGNaiv3} as
\begin{equation} \label{Eq:RecoBGJointFinal}
 \underset{\zbs y}{\text{argmin}}  \Vert \zb A \zb y - \zb w \Vert_2^2 + \Vert \zb D \zb y \Vert_2^2.  
\end{equation}
This least-squares problem is in standard Tikhonov form and thus can be efficiently solved. This optimization problem will be the core of our joined reconstruction algorithm summarized in \ref{Alg1}.

\subsection{Setup the Dictionary}\label{sec:SetupDict}

Having derived an efficient method for the determination of the coefficients $\zb n$ and the tracer distribution $\zb c$ we still need a way to find a good dictionary to describe the background signal. Such a dictionary would ideally be based on a physical model of the background so that $\zb \Phi$ could be derived analytically. However, it is difficult to predict the background signal and its specific spectral fingerprint in practice. We therefore use an alternative approach where the dictionary is determined in a data-driven fashion from a set of $\Theta$ background measurements $\zb u^\text{BG}_\kappa$, $\kappa=1, \dots, \Theta$.
Similar approaches are known from MRI, where the temporal evolution of an image-series is modelled using low rank matrices \cite{Ben2015t2Map, ma2013MRF}. Similarly, low rank matrices are used in audio signal processing to separate the signals from multiple sources \cite{vincent2018audio, 7177939, huang2012singing}.

The required background measurements can be measured over time with the MPI scanner and a free scanner bore. Ideally $\Theta$ is chosen large so that many variations of the background can be tracked. These measurements do not need to be measured in a continuous measurement but it is also possible to use measurements from different scanning sessions. In this way, the dataset can be extended step-by-step and a sufficiently dense sampling of the space of background signals can be achieved. %captures quite accurately the space of background signals. 

With the background measurements at hand, we can setup the background matrix
\begin{align}
    \zb X := \begin{pmatrix}  \zb u^\text{BG}_1 \cdots \zb u^\text{BG}_\Theta \end{pmatrix} \in \mathbb{C}^{M \times \Theta}.
\end{align}
Note that $\zb X$ is not directly suitable as a dictionary since we made the assumption that $\zb \Phi$ is orthogonal, which is not fulfilled by $\zb X$ in general.
We, therefore, propose a rank reduction in combination with an orthogonalization, which can both be achieved by calculating the singular value decomposition of $\zb X$, i.e.
\begin{align}
    \zb X &= \zb U \zb \Sigma \zb V^\adj,
\end{align} %%= \left( U_{m,n} \right)_{m=1,\dots,M;n=1,\dots,M}
where $\zb U 
\in \mathbb{C}^{M\times M}$ and $\zb V \in \mathbb{C}^{\Theta \times \Theta}$ are two unitary matrices and $\zb \Sigma \in \mathbb{R}^{M\times \Theta}$ is a generalized diagonal matrix  containing the singular values $s_1, \dots, s_\Gamma$ on its main diagonal, where $\Gamma < \text{min}(M,\Theta)$ is the rank of $\zb X$. The singular values are stored in non-increasing order such that the most important information is encoded in the first singular values and the associated singular vectors, i.e. the columns of $\zb U$. Since, the background matrix $\zb X$ is setup by measurement of background scans and since the background is changing only slightly over time, the matrix in practice has very similar columns. Furthermore, the matrix $\zb X$ contains noise and in turn one should not consider the true rank of $\zb X$ but truncate the singular values when they fall under a predefined threshold. We propose to chose the $Q$ largest singular values and thus take the first $Q$ columns for the dictionary $\zb \Phi$, i.e.
\begin{align}
  \zb \Phi &= P_{1,M,1,Q}(\zb U) \in \mathbb{C}^{M\times Q}.
\end{align}
Following this line of argumentation, one expects that the first columns of $\zb \Phi$ contribute the most to any given background signals. For the remaining columns, their respective contribution should become smaller along with their associated singular values. To enforce this kind of behaviour in our reconstruction method, we fix the weighting matrix $\zb W$ according to the singular values $s_1, \dots, s_Q$, as follows
\begin{equation*}
    \zb W = \text{diag}(1, \frac{s_1}{s_2}, \dots, \frac{s_1}{s_Q}).
\end{equation*}

The proposed joint background estimation is summarized in algorithm 1. It is defined for a dynamic imaging sequence with $L$ measured frames to underline that some of the operations only need to be performed once. In particular the singular value decomposition of the background data only needs to be performed during the initialization phase of the algorithm.

\begin{algorithm}[t] \label{Alg1}
  \caption{Joint Estimation Algorithm}
  \begin{tabular}{ll}
    Input: & $\zb S \in \mathbb{C}^{M\times N}$, $\zb X \in \mathbb{C}^{M\times \Theta}$, $\zb u_l \in \mathbb{C}^{M}$, $l=1,\dots,L$\\
           & $\zb b^\text{est} \in \mathbb{C}^{M}$, $Q \in \mathbb{N}$, $\lambda,\beta \in \mathbb{R}_+$
  \end{tabular}
  \vspace{5pt}
  \begin{algorithmic}[1] 
    \STATE $\zb U, \zb \Sigma, \zb V^\adj \leftarrow \text{svd}(\zb X)$
    \STATE $\zb \Phi \leftarrow P_{1,M,1,Q}(\zb U)$ 
    \STATE $\zb A \leftarrow \begin{pmatrix} \zb S & \zb \Phi\end{pmatrix}$ 
    \STATE $\zb D \leftarrow \begin{pmatrix} \lambda^{\frac{1}{2}} \zb I_N & \zb 0 \\ \zb 0 & \beta^{\frac{1}{2}} \zb W^{\frac{1}{2}}
    \end{pmatrix}$
    \FOR{$l=1,\dots, L$}
    \STATE $\zb w_l \leftarrow \zb u_l - \zb b^\text{est}$
    \STATE $\zb y_l \leftarrow \underset{\zbs y}{\text{argmin}}  \Vert \zb A \zb y - \zb w_l \Vert_2^2 + \Vert \zb D \zb y \Vert_2^2$
    \STATE $ \zb c_l \leftarrow P_{1,N}( \zb y_l)$
    \STATE $ \zb n_l \leftarrow P_{N+1,N+Q}(\zb y_l)$
    \STATE $ \zb b_l \leftarrow \zb \Phi \zb n_l$
    \ENDFOR
  \end{algorithmic}
  \begin{tabular}{ll} & \\
    Output: & $\zb c_l \in \mathbb{C}^{N}, \zb b_l \in \mathbb{C}^{M}$, $l=1,\dots,L$ 
\end{tabular}
\end{algorithm}

\subsection{Least Squares Solver}

Next, we discuss how to solve the linear optimization problem \eqref{Eq:RecoBGJointFinal} that marks the core of our proposed algorithm\,1. While there exist a multitude of suitable solvers or the least squares problem in MPI, the iterative Kaczmarz method %~\cite{Kaczmarz1937} 
marks the gold-standard in MPI since it converges rapidly \cite{Knopp2010PhysMedBio} thus keeps the entire reconstruction time low. In this work, for simplicity, we only use the Kaczmarz method although any other method is also applicable. In its standard form, Kaczmarz method is only applicable to consistent linear system and not suitable for a least-squares setup. In order to solve
\eqref{Eq:RecoBGJointFinal} with the Kaczmarz method one has to apply a variable substitution $\zb z = \zb D \zb y$ and then solve the optimization problem
\begin{equation} %\label{Eq:RecoBGJointFinal}
 \underset{\zbs z}{\text{argmin}}  \Vert \zb A \zb D^{-1}\zb z - \zb w \Vert_2^2 + \Vert \zb z \Vert_2^2,
\end{equation}
which can be done by applying the Kaczmarz algorithm to the linear system
\begin{align} \label{eq:KaczmarzExtSys}
    \begin{pmatrix} \zb A \zb D^{-1} & \zb I_{M}\end{pmatrix} \begin{pmatrix} \zb z \\ \zb \tau \end{pmatrix} & = \zb w
\end{align}
where $\zb \tau$ is an auxiliary variable that converges to the residual $\zb \tau = \zb w - \zb A \zb D^{-1} \zb z = \zb w - \zb A \zb y$. Once $\zb z$ is calculated by the Kaczmarz algorithm one can determine $\zb y$ by $\zb y = \zb D^{-1} \zb z$. The particle distribution $\zb c$ can then be obtained by taking only the leading $N$ entries of $\zb y$ (c.f. algorithm 1).

\section{Materials and Methods}

\subsection{Experiments}

To evaluate the performance of the proposed algorithm we use the measurement data collected in~\cite{knopp2019correction}. The data was measured with a custom-made human-sized brain scanner~\cite{graser2019human} and allows for qualitative and quantitative analysis of background drifts and background estimation methods. The  data was acquired with a 2D measurement sequence with an excitation field in $x$-direction with an amplitude of \SI{6}{\milli\tesla\per\vacuumpermeability} and frequency of $f_\text{E} \approx\SI{25.599}{\kilo\hertz}$. The excitation field is superposed by a dynamic selection field with a repetition time of approximately \SI{0.5}{\second}, which moves the FFP slowly in $y$-direction. The overall gradient strength was \SI{0.2}{\tesla\per\meter\per\vacuumpermeability} in $y$-direction and half of that value in $x$-direction. The size of the sampled FoV was about \SI{140x140}{\milli\meter\square}. The induced voltage signal was sampled with a rate of $f_\text{ADC} \approx 1.953$~MHz and then block-averaged for data-reduction so that \num{130} line scans orientated in $x$-direction are available in each repetition of the datasets. The number of sampling points per line scan was $\frac{f_\text{ADC}}{f_\text{E}} = 76$ and after applying a real-to-complex discrete Fourier transform a total number of 39 frequency components was available per line scan. All experiments were performed with the MPI tracer perimag (micromod Partikeltechnologie GmbH, Rostock, Germany).

A system matrix was measured using a cubic sized delta sample with a volume of \SI{250}{\micro\litre} (iron mass \SI{4.25}{\milli\gram\of{Fe}}, concentration \SI{17}{\milli\gram\of{Fe}\per\milli\litre}) at \num{28x28} positions in a FoV of size \SI{140x140}{\milli\meter\square} resulting in a voxel size of \SI{5x5}{\milli\meter\square}. In addition \num{5} background scans were performed after measuring each row of the \num{28x28} grid positions such that in total \num{145} background scans were available. Those background scans are used to setup the background matrix $\zb X\in \mathbb{C}^{M\times 145}$. This way of acquiring the background signals during the system matrix measurement was proposed in \cite{knopp2019correction} for background correction of the system matrix and is not specially designed for obtaining an optimal background dictionary. It is also a common method used in the commercial MPI scanners \cite{FrankeHybridMRMPI2016}.

The foreground scans measured during system calibration were used to setup the matrix $\zb S \in \mathbb{C}^{M\times 784}$, where $M$ depends on the frequency selection being used. The system matrix itself was background corrected using the linear interpolation method proposed in \cite{knopp2019correction}. We note that  the tracer concentration used for the system matrix measurement is much larger than the one during the actual experiment and therefore the error of the system matrix is sufficiently small after applying a linearly interpolated background correction.

Two different measurements were performed to evaluate the performance of the proposed background estimation method. In the first measurement a static sample of size \SI{6.3x6.3x6.3}{\milli\cubic\meter} with a low iron mass of \SI{31.25}{\micro\gram\of{Fe}} ( concentration \SI{125}{\micro\gram\of{Fe}\per\milli\litre}) was placed in the center of the scanner and it was measured for about \SI{65}{\second}. The measurement was started before the object placement and finished after the object was removed from the scanner. In turn, right before and after the object placement and removal several background frames are available. We use the mean of 5 frames before the object placement as $\zb b^\text{est}$ and the mean of the 5 frames after the object removal for the linear interpolation method. The 140 frames in-between are used for reconstruction of the particle concentration.

\begin{figure}
\includegraphics[width=1.0\columnwidth]{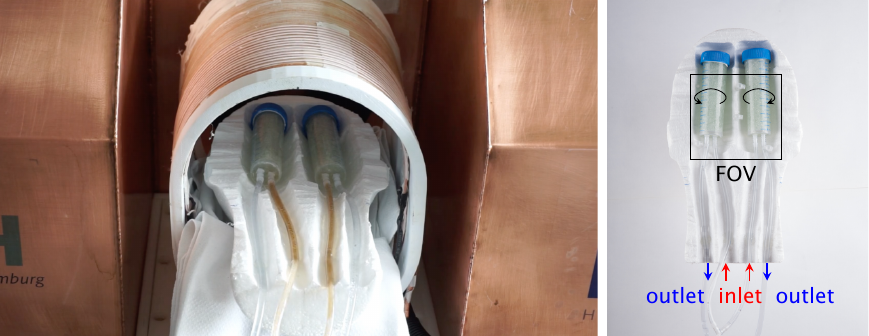}
\caption{Pictures of the dynamic particle phantom that was used to simulate brain perfusion using two tubes that are connected to in-going and out-going hoses. During a continuous circulation with water, a tracer bolus is injected. On the left picture it can be seen how the tracer enters the phantom via the feeding hoses.}
\label{Fig:PhantomDyn}
\end{figure}

In the second experiment a dynamic tracer distribution is considered that would occur in a typical bolus experiment. A human brain is simulated using two tubes that are connected each to an in-going and an out-going hose. Fig.~\ref{Fig:PhantomDyn} shows pictures of the phantom during the experiment which had a duration of \SI{247}{\second} (490 frames). At the beginning of the experiment (about frame 30) a bolus of \SI{100}{\micro\litre} perimag with an iron mass of \SI{850}{\micro\gram\of{Fe}} (concentration \SI{8.5}{\milli\gram\of{Fe}\per\milli\litre})) was injected into the feeding hose of the phantom which remained within the phantom for about 40 frames. 
As in the static sample experiment, background frames are taken directly before (frame 1--5) bolus injection and after (frame 181--185) the tracer has left the phantom. The phantom was not removed from the scanner such that small amounts of tracer could be present within the second background measurement.

\subsection{Image Reconstruction}

In MPI it is common to apply a frequency selection prior to reconstruction \cite{knopp2016online}, which has two potential benefits. First, it allows to remove frequencies, for which no signal is expected such that the residual term takes only the signal carrying frequencies into account. Second, as has been shown in  \cite{them2016sensitivity}, the frequency selection also acts as a background removal method, with the potential downside that valuable data is dropped. We consider a simple band-pass filter by taking into account only frequencies from $f_\text{start}$ until $f_\text{stop}$. The upper frequency is set to $f_\text{stop} = 8f_\text{E}$. Higher frequencies did not contain any measurable MPI signal.
To determine whether it is possible to use additional frequency components, which would usually be dropped because of the background signal drift we consider two different frequency selections. In the first case, which marks the state-of-the art in MPI, we use  $f_\text{start} = 2f_\text{E}$ where $f_\text{E}$ is the excitation frequency. The excitation frequency is commonly excluded in MPI since the signal at $f_\text{E}$ is strongly influenced by the direct feed-through of the excitation field. We name this frequency selection $F_\text{excluding}$. In the second case, named $F_\text{including}$, we take the excitation frequency into account by choosing $f_\text{start} = f_\text{E}$. The number of matrix rows of $\zb S$ and $\zb X$ is $M=6 \cdot 130=780$ for $F_\text{excluding}$ and $M=7\cdot 130=910$ for $F_\text{including}$.

The proposed algorithm has two regularization parameters that need to be appropriately chosen. $\lambda$ is usually chosen relative to the ratio of Frobenius norm of the system matrix and its number of columns $\lambda = \tilde{\lambda} \frac{\text{trace}(\zbs S^\adj \zbs S)}{N}$~\cite{Knopp2010PhysMedBio}, which is also used in this work. Since the concentration in the measurement was chosen rather low, a rather high relative regularization parameters of $\tilde{\lambda} = 1.0$ was chosen. This value was also used in~\cite{knopp2019correction} for the linear background subtraction method. There it was found by visual inspection of the reconstructed images. To reduce the complexity of parameter optimization we keep the regularization parameter $\lambda$ fixed for all reconstructions. In addition, we keep the number of Kaczmarz iterations fixed and use 20 iterations in all reconstructions.

The second regularization parameter $\beta$ is used to adjust the influence of the background term $\beta \Vert \zb n\Vert_2^2$. In order to study the influence of $\beta$ on the reconstruction result we carried out reconstructions with $$\beta_j =  \left(\frac{1}{5}\right)^{j-1}, \quad \text{for}  \quad  j=1,\dots,15.$$
Thus, $\beta$ is selected on a logarithmic scale to sample from a large range of values.

The third parameter to choose is the size $Q$ of the background dictionary $\zb \Phi$. We first analyse the associated singular values of the matrix $\zb X$ and then perform reconstruction with $Q = 1, \dots, 10$ for each of the regularization parameters $\beta_j$. In total we thus perform 150 reconstructions with different values for $\beta$ and $Q$ and analyze the influence of both parameters on the quality of the reconstruction.

To analyze the reconstruction quality we use different quantitative measures. To this end, the reconstructed particle distribution $\zb c$ is first divided into two regions. A \textcolor{green}{$7\times 7$} block of pixels in the center containing the signal is stored in the signal vector $\zb c^\text{signal}$ of length $N^\text{signal}$. All other pixels carry the background signal and are stored in the vector $\zb c^\text{bg}$ of length $N^\text{bg}$. Based on that we calculate the following quantities:
\begin{itemize}
    \item The signal quality is measured by calculating the iron mass 
    \begin{equation}
        m_\text{Fe} = \Delta V \sum_{n=1}^{N^\text{signal}}  c^\text{signal}_n,
    \end{equation}
    where $\Delta V$ is the size of the image pixels. This integrative measure can be compared with the iron mass of the sample placed into the scanner. Using the iron mass has the advantage of being robust against a blurring of the reconstructed particle distribution.
    \item The noise level is measured by calculating 
    \begin{equation}
        \varepsilon_\text{bg} = \frac{1}{c^\text{ref}} \sqrt{ \frac{1}{N^\text{bg}} \sum_{n=1}^{N^\text{bg}}  \left(c^\text{bg}_n\right)^2},
    \end{equation}
    where $c^\text{ref}$ is a reference value of the expected particle distribution taken from a reference reconstruction and calculating $c^\text{ref} = \Vert \zb c^\text{signal} \Vert_\infty$. We note that the same value is taken for all reconstructions and that the purpose of $c^\text{ref}$ is only to report the noise level relative to the signal level.
    \item The third measure is the signal-to-noise ratio of $\zb c$. It is calculated by
    \begin{equation}
        \text{SNR} = \frac{\Vert \zb c^\text{signal} \Vert_\infty}{\frac{1}{\sqrt{N^\text{bg}}}\Vert  \zb c^\text{bg} \Vert_2},
    \end{equation}
    i.e. the maximum signal is relative to the standard deviation of the background signal.
    \item Finally, in order to quantify the spatial resolution of the reconstructed images we calculate the full width at half maximum  (FWHM) of the reconstructed dot in $x$-direction (horizontal direction in the images) through the pixel with the highest intensity. 
\end{itemize}

\section{Results}

\subsection{Background Dictionary Analysis}

\begin{figure}
    \centering
    \includegraphics[width=1.0\columnwidth]{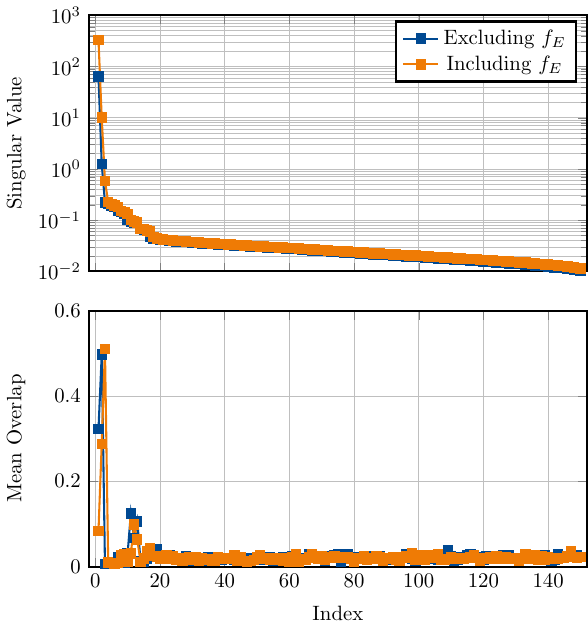}
    \caption{Characteristics of the dictionary $\boldsymbol{\Phi}$ obtained by computing the SVD of the background measurements $\zb X$. The top plot shows the singular values. The bottom plot shows the mean overlap of each basis function with the columns of $\zb S$. 
    }
    \label{fig:svals}
\end{figure}

We start by having a look at the singular values that are computed from the background measurements $\zb X$. They are shown in the top plot of Fig.~\ref{fig:svals} and are computed for both considered frequency selections $F_\text{including}$ and $F_\text{excluding}$. One can see that the singular values drop rapidly in both cases by more than one decade over the first three singular values. This shows that the background signals are very similar. Starting from a certain singular value (5 in case of $F_\text{including}$ and 4 in case of $F_\text{excluding}$), the singular values decay more slowly, which is a typical behavior for an SVD derived from noisy data. Starting from singular value 18, the singular values decay even less indicating that the noise floor is reached. %The first 18 singular values are marked by a gray vertical line. 

Comparing the singular values  for the frequency selections $F_\text{including}$ and $F_\text{excluding}$ one can observe a very similar behavior. When taking a closer look one can see that the singular values for $F_\text{including}$ decay a little bit slower than for $F_\text{excluding}$ indicating that the covered space is a little bit larger. This is not unexpected since the signal at the excitation frequency shows the largest drift in practice.  

Additionally, we computed the mean overlap of each basis function in $\boldsymbol{\Phi}$ with the columns of $\zb S$. We do this by calculating the inner products 
\begin{equation*}
 \alpha_{n,\gamma} = \frac{\langle \zb \phi_\gamma, \zb s_n \rangle_2}{\Vert \zb  \phi_\gamma \Vert_2 \Vert \zb s_n \Vert_2},  \quad n=1,\dots,N, \; \gamma=1,\dots,\Gamma
\end{equation*}
 where $\phi_\gamma$ and $\zb s_n$ are columns of $\zb \Phi$ and $\zb S$, respectively. As an integrative measure we then calculate $\bar{\alpha}_\gamma = \frac{1}{N}\sum_{n=1}^{N} \vert \alpha_{n,\gamma}\vert$.
As can be seen in the bottom plot of Fig.~\ref{fig:svals}, the first three basis functions have a significant overlap with the space spanned by the system matrix columns. For the remaining basis functions, the overlap quickly decreases with most of the values being smaller than 0.1. The overlap of the first basis functions illustrates the need for a proper treatment of background signals when performing MPI image reconstruction. In fact, a negligible overlap would lead to the situation that background signals cannot be falsely attributed to particle signals. In this case, even a standard reconstruction based on \eqref{Eq:RecoStd} should be sufficient to obtain images without background artifacts.

\subsection{Parameter Optimization}

\begin{figure}[t]
    \includegraphics[width=\columnwidth]{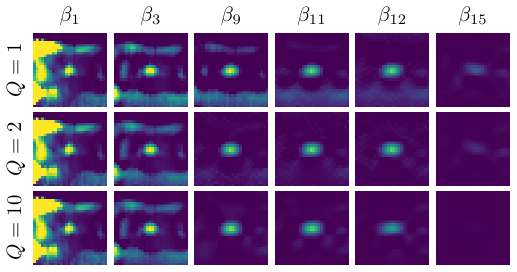}
    \caption{Reconstruction results using the joint estimation algorithm for frame $l=121$ of the static dot phantom experiment including the excitation frequency. Shown are results for $\beta_j$, $j=1,3,9,11,12,15$ and $Q=1,2,10$}
    \label{fig:RecoParamOptim}
\end{figure}

\begin{figure}[t]
    \includegraphics[width=\columnwidth]{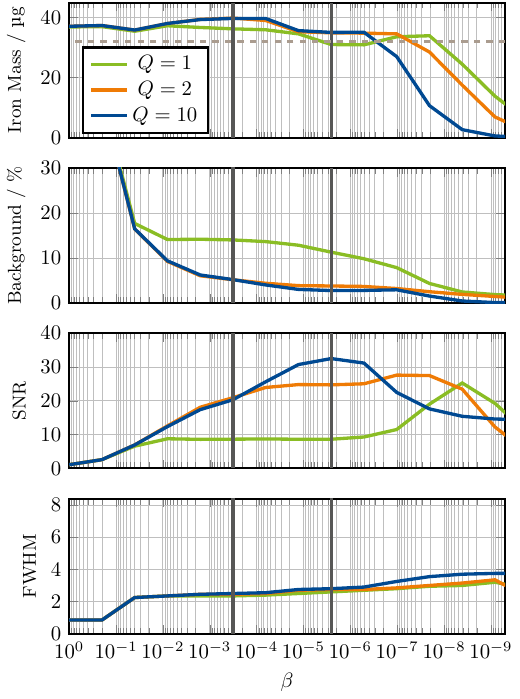}
    \caption{Quantitative measures of the image quality of the proposed reconstruction algorithm for frame $l=121$ of the static phantom experiment including the excitation frequency. Shown are the iron mass $m_\text{Fe}$ (first plot), the background $\varepsilon_\text{bg}$ (second plot), the SNR (third plot), and the FWHM (last plot) for $Q=1,2,10$ and $\beta \in [(\frac{1}{5})^{14}, 1]$. The selected $\beta$-values are indicated by vertical dark gray lines. In the first plot the expected iron mass is drawn as a dashed brown horizontal line.}
    \label{fig:ParamOptim}
\end{figure}

After analyzing the singular values we next consider the reconstruction results for the static phantom experiment and investigate the influence of the parameters $\beta$ and $Q$ on the image quality. The analysis is performed for a single frame ($l=121$) at the end of the measurement, when the background has already drifted considerably compared to the initial background frame $\zb b^\text{est}$. We only show the parameter optimization results for the data where the excitation frequency is included since the findings are similar for the case where the excitation frequency was excluded.

Reconstruction results for a subset of parameters are shown in Fig.~\ref{fig:RecoParamOptim}. In the images one can see that for large $\beta$ strong background artifacts are present. In fact, for $\beta_1$ the influence of the background estimation term is very small and thus the image looks very similar to the one obtained with static background subtraction, which is shown in the forth column and first row in the lower part of Fig.~\ref{fig:RecoStatic1}. When decreasing $\beta$ the deviation from the initial background estimate $\zb b_\text{est}$ have a smaller cost in the optimization functional \eqref{Eq:RecoBGNaiv1} and in turn a more accurate background signal is estimated, which can be inferred from the reduced artifact level in the images. When decreasing $\beta$ even further one can see that at some point the spatial resolution of the reconstructed dot decreases up to the point that the entire signal vanishes. All these observations can also be verified in the quantitative measures shown in Fig.~\ref{fig:ParamOptim}. Here, one additionally sees that the reconstructed iron mass remains in a similar range as the true iron mass until a certain value of $\beta$ after which the iron mass is underestimated. 

When taking a look at the influence of $Q$ one can first observe that the asymptotic behavior for $\beta$ is similar for different $Q$. But the effect that the particle distribution vanishes happens the earlier the larger the dictionary $Q$ is. In the images shown in Fig.~\ref{fig:RecoParamOptim} one can see an artifact remaining for $Q=1$ and most of the $\beta$ values. This artifact is removed when either taking a very small value for $\beta$ or when increasing $Q$ to 2.

By visual inspection of the image data and the derived quantitative values shown in Fig.~\ref{fig:ParamOptim}, we selected a value of $\beta = \left(\frac{1}{5}\right)^{8}$ and a dictionary size of $Q=10$ for further reconstructions of the data. For further comparison the same reconstructions were also performed with a value of $\beta = \left(\frac{1}{5}\right)^{5}$.

\subsection{Static Experiment}

\begin{figure}

    \includegraphics[width=\columnwidth]{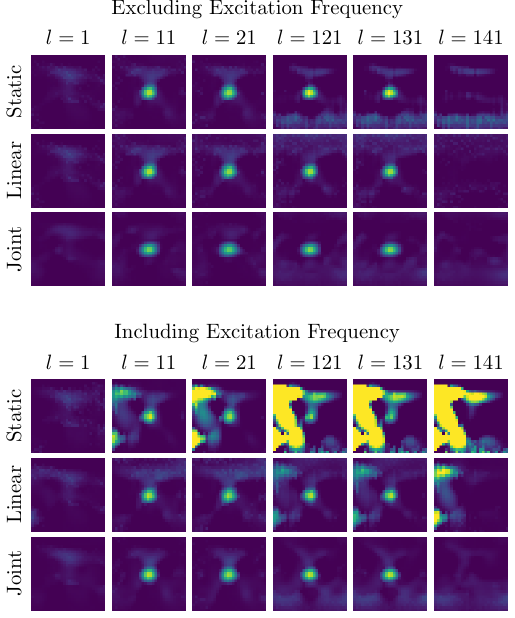}
    \caption{Reconstruction results for the static dot phantom placed in the scanner center between frames $l=11$ and $l=131$. The images were
    reconstructed with (bottom) and without (top) the excitation frequency. In both parts selected frames of the time series reconstructed with static background subtraction (first row), linearly interpolated background subtraction (second row) and the proposed joint background and particle distribution estimation (third row) are shown. The later were calculated with $\beta=\left(\frac{1}{5}\right)^{5}$ and $Q=10$ for both frequency selection schemes.}
    \label{fig:RecoStatic1}
\end{figure}

\begin{figure*}

    \includegraphics[width=\textwidth]{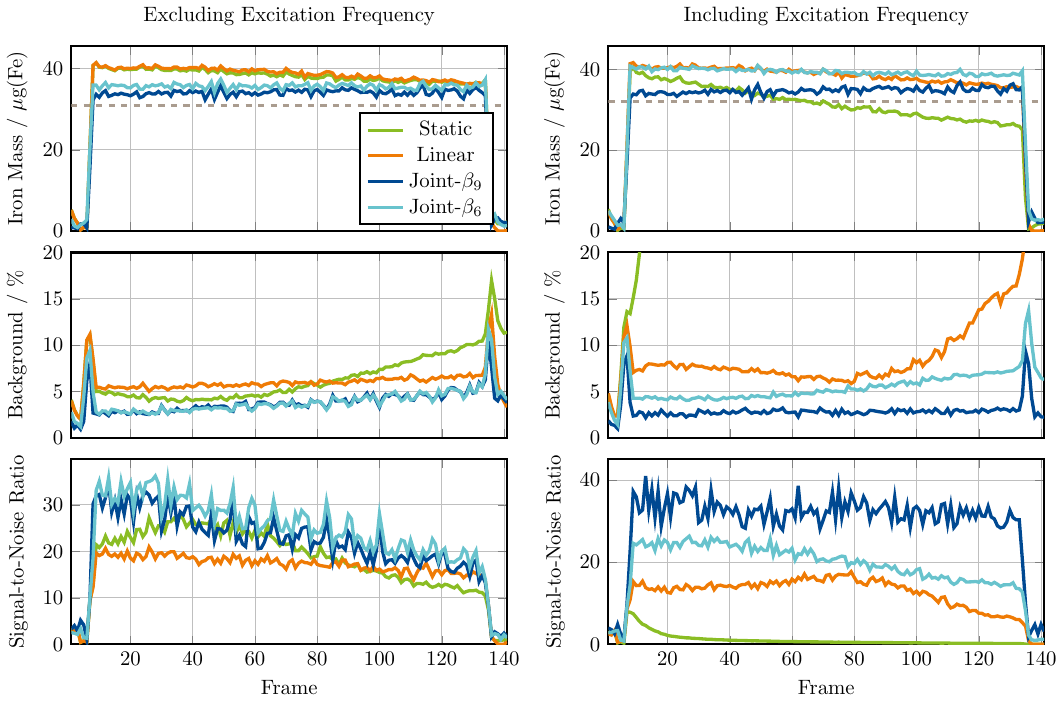}
    \caption{Iron mass, background level and SNR
    of the reconstructed data shown in Fig.~\ref{fig:RecoStatic1}
    for the static and linearly interpolated background subtraction as well as for the joint estimation approach. In case of the joint estimation all quantities were computed both for the optimized $\beta$-value of $\beta_9=\left(\frac{1}{5}\right)^8$ and for a larger value of $\beta_6=\left(\frac{1}{5}\right)^5$.
    In case of the iron mass, the true iron mass (31~$\mu$g(Fe)) is shown as a horizontal dashed brown line.}
    \label{fig:SNR}
\end{figure*}

After tuning the background-related reconstruction parameters we next consider the entire time series of the experiment and compare the proposed joint estimation algorithm with the static and the linearly interpolated background subtraction. 
As one can see in Fig.~\ref{fig:RecoStatic1}, the static background subtraction is not capable of preventing background artifacts over time since the background signal is drifting. Thus the artifacts become stronger than the actual signal. The artifacts are much stronger in the case that the excitation frequency is included. The linear interpolation of the background substantially reduces the background artifacts. While  only slight artifacts are present in the results excluding the excitation frequency the results including the excitation frequency still show artifacts with similar strength as the particle signal itself (see frame $l=131$). In contrast, the joint estimation approach is capable of suppressing the artifact even in the challenging case that the excitation frequency is included. The reconstructed dot is slightly more blurred but the overall image quality is the same or better than in the case of the static or linearly interpolated background subtraction.

Quantitative measures of the reconstructed images are shown in Fig.~\ref{fig:SNR}. One can clearly see that the joint estimation approach outperforms the other methods with respect to the background suppression and the overall SNR. One additional advantage is outlined in the iron mass plot. One can clearly see that the static and the linearly interpolated background subtraction approach show a non-constant progression of the estimated iron content, which can only be caused by the drift of the background signal. In contrast, the joint estimation approach does not show a systematic drift in the iron mass over time. Additionally, we note that all methods systematically overestimate the iron mass. Part of this bias can certainly be explained by errors in the manufacturing of the phantom. Nevertheless, it can also be observed that the joint reconstructions yield results with an iron mass that is overall closer to the reference value of 31.25~$\mu$g(Fe).

\subsection{Dynamic Experiment}

\begin{figure*}
    \centering
    % \vspace{-15pt}
    \includegraphics[width=\textwidth]{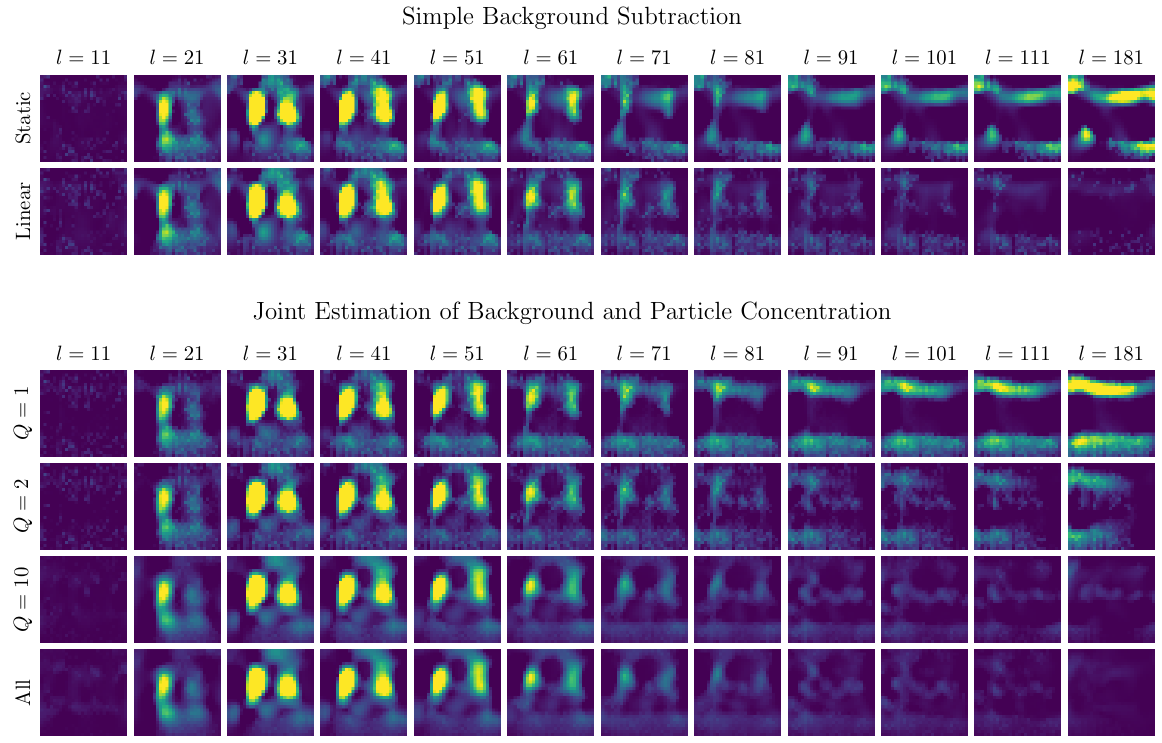}
    \caption{Reconstruction results for the dynamic bolus experiment using frequency selection $F_\text{excluding}$. In the upper part, the results for the static and a linearly interpolated background subtraction are shown. Below the results for the joint estimation approach for varying $Q\in \{1,2,10\}$ and $\beta=\left(\frac{1}{5}\right)^5$ are shown. In the last row, the results are shown for the case that the entire matrix $\zb U$ is used as the dictionary $\zb \Phi$. All images are equally windowed and focused on the smaller particle concentrations (25\,\% of the maximum value in the time series where the background was corrected by linear interpolation)}
    \label{fig:recoDyn}
\end{figure*}

The results of the dynamic bolus experiment are shown in Fig.~\ref{fig:recoDyn}. Since the results for the different frequency selections look very similar only the results for frequency selection $F_\text{excluding}$ are shown. It can be seen that the static background subtraction method shows a plausible reconstruction result where the concentration in the two tubes first increases (frame $l=21-41$) and then decreases. Starting from frame $l=91$ the concentration no longer decreases and instead background artifacts appear in regions, where the phantom does not contain any particles (see Fig.~\ref{Fig:PhantomDyn}). The linear background interpolation method is able to mostly remove this artifact and provide a plausible result. The results for the proposed background estimation method are shown in the lower part of Fig~\ref{fig:recoDyn}.  It can be seen that the background artifact is still present for a dictionary of size $Q=1$. However, when the dictionary is enlarged, the artifact is also suppressed, with the best results for $Q=10$. A further positive effect of the increase of the dictionary is that the image noise is substantially reduced. We note that a further increase of the dictionary taking into account all singular vectors did not change the result substantially but only slightly smoothed the particle signal.

\section{Discussion}

In the present work we introduced a new method for background estimation and removal in magnetic particle imaging. In both the static and the dynamic experiment the method outperformed the static background subtraction method and performed similar or better than the linearly interpolated background subtraction method. Beyond image quality, the proposed joint estimation approach has the important advantage over the other methods that no complicated imaging protocol, with measurements right before and after the measurement, is strictly required. This is a strong advantage in particular for long measurements where the background evolves non-linearly over time.

A main difference to static and linearly interpolated background subtraction is that the proposed method does not make any strong assumptions on the temporal evolution of the background signal. Instead the background is estimated jointly with the tracer concentration on a frame-by-frame basis. As a consequence, even non-linear temporal evolution of the background can be captured, as long as the background signal is well described by the used dictionary. Nevertheless, we note that acquiring a background measurement right before the measurement can still be beneficial since this reduces the magnitude of the signal drifts which need to be estimated. How important a decent background estimation is depends highly on the strength of the baseline background signal and therefore cannot be easily generalized. In our case it was not possible to reconstruct a satisfying particle distribution for the static experiment when setting the background estimate $\zb b_\text{est}$ to zero. It was possible, however, to use background estimates from previous measurements without strong degradation of the image quality.

The key idea of the algorithm is to separate the background signals from the particle signals by considering the space spanned by the particle signals (columns of $\zb S$) and the space spanned by the background signals (columns of  $\zb \Phi$). The later is derived in a data-driven manner from a set of background scans that can be obtained e.g. during a system matrix acquisition. How well this approach works, depends highly on the spaces having a sufficiently small intersection. For the scanner being used, the spaces were different enough to perform a separation of particle signal and background signal. We note however that further experiments across different MPI scanners are necessary to study background signals in more detail.

One important aspect of our study was to investigate the influence of the regularization parameter $\beta$. We have seen that the influence of $\beta$ on the background removal was the smaller the larger $\beta$ was. This is plausible from a mathematical point of view when looking at the optimization problem \eqref{Eq:Deriv2} that forms the basis of the algorithm.
In the limit $\beta \rightarrow \infty$ the regularization term $\Vert \zb \Phi \zb n \Vert_2$ would dominate the other terms and in turn $\zb n = \zb 0$ would be the solution of the minimization problem so that $\zb c$ remains untouched. This solution thus converges for $\beta \rightarrow \infty$ to the solution obtained by static background subtraction. 

The smaller $\beta$, the less is $\zb n$ restricted. This can lead to particle signal being falsely identified as background. In turn this puts a bias on the reconstructed particle concentration, such that one has to trade off and select an appropriate regularization value $\beta$ with strong background suppression but only marginal bias on the estimation of the particle distribution. Our quantitative analysis has shown that the choice of $\beta$ is rather robust yielding satisfactory results for a large range of values. The influence of the size of the background dictionary $Q$ depends highly on the specific setup and the statistics of the background signal. In our case, $Q=10$ yielded satisfying results for all experiments.

Our proposed approach for selecting $\lambda, \beta$ and $Q$ is to first chose a large enough value for $Q$ and keep this parameter static. Then one sets $\beta=\infty$ and optimizes $\lambda$ to obtain a good image quality as one would do in a regular Tikhonov regularized MPI reconstruction. In the last step $\beta$ is optimized to suppress the background artifacts. While this procedure does not necessary yield the optimum parameters it has the advantage that the parameters are optimized sequentially, which is much easier to tackle than a simultaneous  optimization of all regularization parameters. However, we note that a more sophisticated parameter optimization approach could potentially allow for choosing smaller $\lambda$ in case that the background signal is predicted well. This could then even increase the spatial resolution of the reconstructed images.

In this work, we did not put a focus on studying the background signal itself over longer time periods to gather a better statistical model of the scanner. For a routine application of the algorithm we advice to establish a database of background scans, which is continuously analysed and extended. In this way the background dictionary gets more and more expressive in case that the background of the MPI system changes over time and $Q$ is chosen large enough. Whenever system components are changed -- for example an exchange of the excitation coils -- new background scans should be taken.

Another aspect to consider when using a background dictionary is that the dictionary was built for a specific imaging sequence, similar to the calibration of the system matrix. It is an open question whether it is possible to build a dictionary that can handle changes in experimental parameters such as the selection field gradient and the excitation field strength, or whether a separate background dictionary is required for each parameter setting.

Finally we want to discuss the computational aspects of the proposed reconstruction approach. Since we use a linear estimator for the background signal, the algorithmic complexity in terms of the ${\cal O}$ notation remains to be ${\cal O}(MNI)$, where $I$ is the number of Kaczmarz iterations, independent of whether the background is estimated or not. The size of the linear system to be solved is only marginally increased since one chooses $Q \ll N$ in practice. Finally, we did not see an increase of the necessary number of Kaczmarz iterations such that we conclude that the overall reconstruction time is only marginally increased for the proposed algorithm \ref{Alg1}.

The efficiency of our algorithm is an advantage over the algorithm proposed in \cite{straub2018joint}, which also performs a joint estimation of the particle distribution and the background signal but leads to a 400-fold increase in computation time, compared to standard reconstruction without background estimation, due to the use of a Newton-type solver. In appendix A %\ref{sec:linearityOfJointBG} 
we show that the algorithm proposed in~\cite{straub2018joint} can also be efficiently solved since the underlying optimization problem can be written as a least-squares problem. We note, however, that~\cite{straub2018joint} requires a special imaging sequence while our approach can be applied in a more general setting. Both algorithms are similar in the way that they constrain the background signal. The purpose of the appendix is to derive the common structure of both background estimation approaches.

%\appendix

\subsection*{Appendix A: Comparison with the Joint BG Estimation Approach proposed in~\cite{straub2018joint}} \label{sec:linearityOfJointBG}

In this appendix our method is compared with the background estimation approach proposed in~\cite{straub2018joint}. We show that both approaches have a common structure and that the method proposed in~\cite{straub2018joint} can be reformulated as a least squares problem.

The idea of the method is to apply a sequence, where the FoV is slightly shifted from frame to frame. We can express this with two shifting operators $\Delta^1$, $\Delta^2$ and the associate measurements $\zb u^1$, $\zb u^2$. Then, the authors in~\cite{straub2018joint} proposed to solve the optimization problem
\begin{align} \label{Eq:RecoBGJointSeq}
 \underset{\zbs c, \zbs b}{\text{argmin}}  \sum_{q=1}^{2}\Vert \zb S \Delta^q (\zb c) - \zb u^q + \zb b \Vert_2^2 &+ \lambda \Vert \zb c \Vert_2^2  + \beta \Vert \zb b - \zb b^\text{est} \Vert_2^2
\end{align}
Here, we can already see various similarities to our approach\,\eqref{Eq:RecoBGNaiv1}. Both approaches have the same background regularization term $\Vert \zb b - \zb b^\text{est} \Vert_2^2$ and the same regularization parameters $\lambda$ and $\beta$. The difference is that \eqref{Eq:RecoBGJointSeq}
includes two subsequent frames and assumes that they have the same background and the same particle concentration. This restricts the background to a low dimensional space since the imaging equation for both frames need to be fulfilled. In contrast, our approach operates on a single frame only and instead restricts the space of the background signals based-on an a-priori chosen dictionary.

We next show that \eqref{Eq:RecoBGJointSeq} is a least-squares problem that can be efficiently solved. To this end, we first note that the shifting operator $\Delta^q(\cdot)$ can be expressed as a matrix-vector multiplication
\begin{align}
  \Delta^q(\zb c) & = \zb H^q \zb c  \quad \text{for} \quad q=1,2,
\end{align}
where $\zb H^q$ is a permutation matrix having a diagonal structure. One may need to zero pad $\zb c$ for proper handling of the boundary pixels.
We then introduce shifted versions of the system matrix $\zb S^q  = \zb S \zb H^q$ such that we can write
\begin{equation*}
\zb S \Delta^q (\zb c) = \zb S \zb H^q \zb c = \zb S^q \zb c.
\end{equation*}
Inserting this into \eqref{Eq:RecoBGJointSeq} yields
\begin{equation*}
 \underset{\zbs c, \zbs b}{\text{argmin}}  \sum_{q=1}^{2}\Vert \zb S^q \zb c - \zb u^q + \zb b \Vert_2^2 + \lambda \Vert \zb c \Vert_2^2 + \beta \Vert \zb b - \zb b^\text{est} \Vert_2^2.  
\end{equation*}
We then move the background reference $\zb b^\text{est}$ into the residual term by variable substitution:
\begin{equation*}
 \underset{\zbs c, \zbs b}{\text{argmin}}  \sum_{q=1}^{2}\Vert \zb S^q \zb c - \zb u^q + \zb b + \zb b^\text{est} \Vert_2^2 + \lambda \Vert \zb c \Vert_2^2 + \beta \Vert \zb b \Vert_2^2,
\end{equation*}
Then, we derive $$\zb S^q \zb c - \zb u^q + \zb b + \zb b^\text{est} = \begin{pmatrix} \zb S^q & \zb I_M \end{pmatrix} \begin{pmatrix} \zb c \\ \zb b \end{pmatrix} -\zb u^q + \zb b^\text{est}$$ and stack the two residual norms as well as the two norms acting on $\zb c$ and $\zb b$ yielding
\begin{equation*} 
 \underset{\zbs c, \zbs b}{\text{argmin}} 
\left\Vert \begin{pmatrix} \zb S^1 & \zb I_M \\ \zb S^2 & \zb I_M  \end{pmatrix} \begin{pmatrix} \zb c \\ \zb b \end{pmatrix} - \begin{pmatrix}\zb u^1 - \zb b^\text{est} \\ \zb u^2 - \zb b^\text{est} \end{pmatrix}\right\Vert_2^2 +
 \left\Vert \begin{pmatrix} \lambda \zb c \\ \beta \zb b\end{pmatrix} \right\Vert_2^2.
\end{equation*}
If we then define
\begin{align*} 
\zb A &:= \begin{pmatrix} \zb S^1 & \zb I_M \\ \zb S^2 & \zb I_M  \end{pmatrix}, \quad
\zb x := \begin{pmatrix} \zb c \\ \zb b \end{pmatrix}, \\
\zb y &:= \begin{pmatrix}\zb u^1 - \zb b^\text{est}\\ \zb u^2 - \zb b^\text{est} \end{pmatrix}, \quad     \zb D := \begin{pmatrix} \lambda^{\frac{1}{2}} \zb I_N & \zb 0 \\ \zb 0 & \beta^{\frac{1}{2}} \zb I_Q
    \end{pmatrix}
\end{align*}
we end up with
\begin{equation} \label{Eq:finalBGEstStraub}
 \underset{\zbs x}{\text{argmin}} 
\left\Vert \zb A \zb x - \zb y \right\Vert_2^2 +
 \left\Vert \zb D\zb x  \right\Vert_2^2,  
\end{equation}
which is a common Tikhonov regularized  least-squares problem that can be efficiently solved. In comparison with our proposed background estimation approach \eqref{Eq:RecoBGJointFinal} mainly the matrix $\zb A$ and the vector $\zb y$ differ in \eqref{Eq:finalBGEstStraub}. Moreover, the regularization function for the background signals is a classical $\ell_2$-regularization, which can be viewed as a special case of the weighted $\ell_2$-regularization used in this work.

\subsection*{Appendix B: Code and Examples}
To make our algorithm accessible for other researchers we integrated it into the open source MPI reconstruction package \textit{MPIReco.jl}~\cite{MPIReco}, which can be accessed under

\url{https://github.com/MagneticParticleImaging/MPIReco.jl}. 

\noindent An example script that uses the data from the static sample experiment is provided in the repository

\url{https://github.com/IBIResearch/DictionaryBasedBackgroundEstimation}

\noindent It automatically downloads the data and generates a subset of Fig.~\ref{fig:RecoStatic1} (DOI of example: 10.5281/zenodo.4972554, DOI of data: 10.5281/zenodo.4972123).

\bibliographystyle{IEEEtran}

%\bibliography{ref}
%\printbibliography

% Generated by IEEEtran.bst, version: 1.14 (2015/08/26)

\end{document}